\documentclass[preprint,12pt]{elsarticle}
\usepackage{amsmath}
\usepackage{amssymb}
\usepackage[dvipdfm,colorlinks]{hyperref}
\usepackage{multirow}

\biboptions{compress}

\begin{document}

\begin{frontmatter}

\title{Secure multiparty quantum computation based on \\ bit commitment}

%\author[label1]{}
\author{Min Liang}\ead{liangmin07@mails.ucas.ac.cn}
\address{Data Communication Science and Technology Research Institute, Beijing 100191, China}

\begin{abstract}
This paper studies secure multiparty quantum computation (SMQC) without nonlocal measurements. Firstly, this task is reduced to secure two-party quantum computation of nonlocal controlled-NOT (NL-CNOT) gate. Then, in the passive adversaries model, the secure computation of NL-CNOT is reduced to bit commitment. Thus, a SMQC scheme can be constructed based on bit commitment. This scheme does not depend on trusted third party, and is secure in the passive adversaries model. It is also pointed out that a vulnerability exists in any secure two-party quantum computation protocol of NL-CNOT gate.
\end{abstract}

\begin{keyword}
Quantum cryptography \sep secure multiparty computation \sep bit commitment \sep quantum circuit \sep passive adversaries
\end{keyword}

\end{frontmatter}

% \linenumbers
\section{Introduction}
Secure multiparty computation is a fundamental cryptographic primitive in modern cryptography. It focuses on the studies of secure computation among the players that do not trust each other. In quantum cryptography, it is also studied extensively as secure multiparty quantum computation (SMQC). The SMQC has been studied from two aspects: 1) the evaluation of classical function with quantum protocol, and 2) the evaluation of quantum transformation.

Lo \cite{lo1997a} studied one-sided two party computation of classical function, and proved that the task cannot be realized securely with quantum protocols. Later, Refs. \cite{colbeck2007,buhrman2012} strengthened this impossibility result: two-sided secure two-party computation of classical function is also impossible with quantum protocols.

Secure multiparty computation of quantum circuit is also studied \cite{crepeau2002,benor2006,unruh2010,dupuis2010,dupuis2012}. Ref. \cite{crepeau2002} presented a verifiable quantum secret sharing protocol, based on which constructed a SMQC scheme. This construction of SMQC can tolerate $\lfloor\frac{n-1}{6}\rfloor$ cheaters among $n$ players. This threshold was improved to	 $\lfloor\frac{n-1}{4}\rfloor$ in \cite{benor2006}.
Dupuis et al. studied secure two-party quantum computation, and proposed a two-party protocol for secure evaluation of unitaries against specious adversaries \cite{dupuis2010}, later the protocol was improved to securely compute any quantum operation against active adversaries \cite{dupuis2012}.

Recently, Ref. \cite{liang2013} presented a quantum fully homomorphic encryption (QFHE) scheme, and described a SMQC scheme of unitaries with trusted third party (TTP) based on QFHE scheme.

This paper studies SMQC without nonlocal measurements. The SMQC task is reduced to secure two-party quantum computation of nonlocal CNOT (NL-CNOT), and then the protocol presented in \cite{liang2013} can be simplified. Then the secure computation protocol of NL-CNOT is reduced to bit commitment.

\section{Preliminaries}
Firstly, the model of SMQC studied in this paper is introduced here. Suppose there are $n$ parties who jointly perform a computational task of quantum circuit. The input of the circuit is a $m$-qubit state. The $n$ parties are denoted as $P_1,\cdots,P_n$, whose inputs have $k_1,\cdots,k_n$ qubits
($m=\sum_{i=1}^n k_i$), respectively. Through their local quantum computation and mutual communication, they accomplish the computational task, and each party can obtain the desired result. The quantum circuits are limited to be those containing no nonlocal measurements.

In this model, any two of the $n$ parties are mutually distrusted, and each party may be dishonest. According to the possible cheating behaviors of dishonest participants, the SMQC can have the different models, such as passive adversaries model, active adversaries model, or else. Here, passive adversaries model refers that the dishonest parties merely gather information during the execution of SMQC protocol, and the active adversaries model refers that the dishonest parties take active steps to disrupt the execution of multiparty protocol. In any multiparty protocol, each party may change its local input before even entering the execution of the protocol. This is also unavoidable when the parties utilize a trusted party. So, this is not considered a breach of security \cite{goldreich2008}.

Ref. \cite{liang2013} constructed a SMQC of unitaries with TTP based on QFHE. The scheme is described as follows: each party ($P_1,\cdots,$ or $P_n$) preshares a secret key with TTP through quantum key distribution (QKD); Then each party encrypts his data with his secret key, and sends the ciphertext to TTP; Using the preshared secret key, the TTP performs the quantum circuit on the encrypted data according to QFHE scheme; The TTP sends back the ouputs to the corresponding parties, and then each party can decrypt the received state and obtain the desired result. This scheme is secure not only in the passive adversaries model but also in the active adversaries model.

For some protocol $\sigma$ and some protocol $\pi$, we denote by $\sigma^\pi$ the protocol where $\sigma$ invokes instances of $\pi$.

{\bf Quantum universal composition (UC) theorem \cite{unruh2010}:} Let $\pi$, $\rho$ and $\sigma$ be quantum polynomial time protocols. Assume that $\pi$ quantum-UC-emulates $\rho$. Then $\sigma^\pi$ quantum-UC-emulates $\sigma^\rho$.

This theorem holds for both computational security and statistical security (information-theoretical security or perfect security). Quantum UC theorem ensures that if the quantum protocol $\pi$ can securely realize a functionality $\mathcal{F}$, then the protocol $\sigma^\pi$
can securely realize a functionality $\sigma^\mathcal{F}$.

\section{Multiparty quantum computation with trusted third party}\label{sec3}
This section introduces the reduction from multiparty quantum computation to two-party quantum computation of NL-CNOT.
Any unitary transformation can be decomposed into some CNOT and single-qubit transformations, so any unitary transformation $U$ can be expressed by a quantum circuit $C_U$ that consists of only CNOT and single-qubit gates. We consider $n$ parties $P_1,\cdots,P_n$ participate in the joint computation of the quantum circuit $C_U$. Each of them $P_i$ has $k_i$ qubits as the inputs of the circuit, respectively. The circuit $C_U$ has $m=\sum_{i=1}^n k_i$ qubits as its input.

The $m$-qubit input of the circuit comes from all the $n$ parties. Because the CNOT gate is performed on two qubits, the CNOT gates in the circuit can be classified into two kinds: (1) local CNOT, which acts on the two qubits belonging to the same party, and (2) NL-CNOT, which acts on the two qubits belonging to two different parties.

\begin{figure}[htp!]
\begin{center}
\includegraphics[width=8cm]{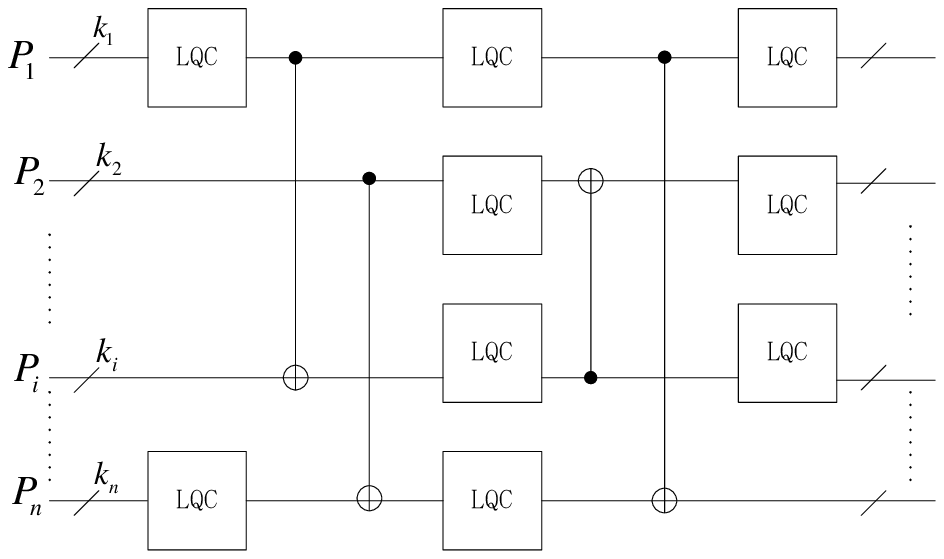}% Here is how to import EPS art
\end{center}
\caption{\label{fig1} Quantum circuit to be computed jointly by $n$ parties. It can be seen as a combination of some NL-CNOT gates and local quantum circuits (LQCs).}
\end{figure}

For an arbitrary party $P_i$, there are many single-qubit gates performing on his $k_i$-qubit input in the circuit. According to the computing sequence, these single-qubit gates are split into some small-scale quantum circuits by the NL-CNOT gates in the circuit. The small-scale quantum circuit acts only on his $k_i$ qubits, so it is called local quantum circuit (LQC). Thus, the joint quantum circuit $C_U$ can be seen as a combination of some NL-CNOT gates and LQCs. For example, any joint quantum circuit $C_U$ can be expressed similarly to this form in Figure \ref{fig1}.

Because these LQCs are performed on their local qubits, the joint computation of the whole quantum circuit $C_U$ only requires the joint computation of every NL-CNOT gate. According to quantum UC theorem \cite{unruh2010}, if the two-party computation of NL-CNOT is secure, then there exists a SMQC protocol for the unitary quantum circuit $C_U$. Thus, the SMQC of unitary circuit is reduced to two-party quantum computation of NL-CNOT.

In general quantum circuit, there may be some quantum measurements. The measurements may be local or nonlocal. The local measurement means that it acts simultaneously on the qubits belonging to the same party. If the quantum circuit contains no nonlocal measurement, the SMQC of quantum circuit can also be reduced to two-party quantum computation of NL-CNOT.

In order to implement ideal NL-CNOT functionality by secure two-party protocol, we can still adopt the TTP in the similar way to the SMQC presented in \cite{liang2013}. The difference is that it is unnecessary to employ the QFHE scheme, while the QHE scheme of CNOT is sufficient here. The detail is as follows. Suppose Alice and Bob intend to compute a NL-CNOT gate, they encrypt their qubits ($|\alpha\rangle$ and $|\beta\rangle$) using the secret key preshared with TTP separately, and send the two qubits to the TTP, then the TTP performs some quantum operation and sends back the result, finally Alice and Bob decrypt the received qubits separately and obtain the desired state $CNOT(|\alpha\rangle\otimes|\beta\rangle)$.

The $n$ parties intend to jointly compute a quantum circuit consists of some NL-CNOTs and LQCs (the LQC may contain local measurements). The LQCs can be computed locally by themselves. Once they have to jointly compute a NL-CNOT gate, they call the TTP. The call procedure is described as above. Because the secret key is needed during the call of TTP, they have to preshare some secret key through QKD, which is unconditionally secure. In this way, with the help of TTP, we can construct a SMQC protocol for quantum circuit without nonlocal measurements. This protocol is just an improvement of the scheme presented in Ref. \cite{liang2013}. In this improved scheme, one computation of NL-CNOT needs one call of TTP. So it is an interactive scheme, and each party does not interact with another party. Moreover, the rounds of interaction depend on the number of NL-CNOT gates in the quantum circuit.

\section{Two-party quantum computation of NL-CNOT}\label{sec4}
It can be concluded from the previous section that, NL-CNOT plays a fundamental role in multiparty quantum computation. Here the secure two-party computation of NL-CNOT will be investigated as the key point. It will be reduced to bit commitment in the passive adversaries model and then a secure two-party quantum computation protocol of NL-CNOT will be proposed.

Denote the four Bell states as: $|B_{xz}\rangle=\frac{1}{\sqrt{2}}(|0x\rangle+(-1)^z|1\bar{x}\rangle)$, where $x,z\in\{0,1\}$, and $\bar{x}=1-x$. Define quantum entanglement state
$$|\chi\rangle=\frac{1}{\sqrt{2}}(|00\rangle+|11\rangle)|00\rangle+\frac{1}{\sqrt{2}}(|01\rangle+|10\rangle)|11\rangle.$$
This quantum state $|\chi\rangle$ can be prepared using two copies of Bell state $|B_{00}\rangle$ and a CNOT gate, e.g. $|\chi\rangle=(I\otimes CNOT \otimes I)B_{00}\otimes B_{00}$.

\begin{figure}[htp!]
\begin{center}
\includegraphics[width=8cm]{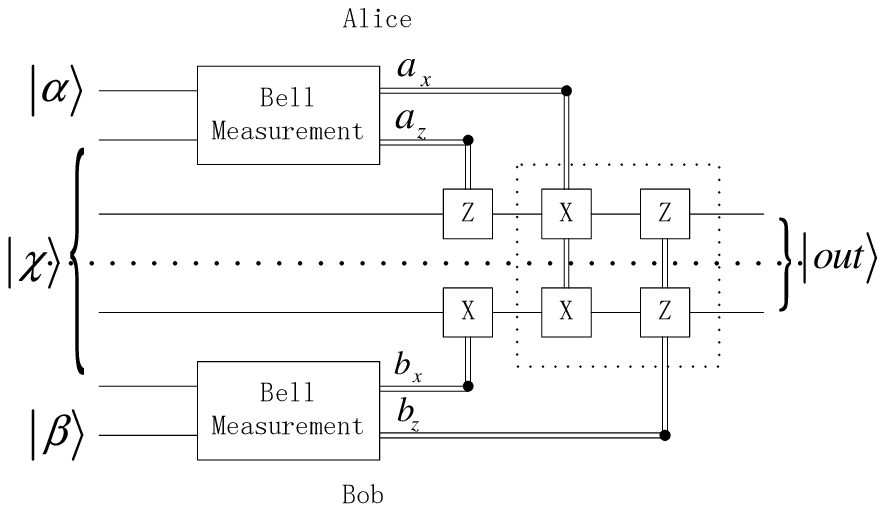}% Here is how to import EPS art
\end{center}
\caption{\label{fig2} The remote implementation of NL-CNOT. The above/below of the dotted line expresses the local quantum operations of Alice/Bob. The dotted rectangle contains nonlocal computation.}
\end{figure}

According to the implementation introduced in Refs.\cite{dupuis2010,gottesman1999}, NL-CNOT can be implemented as Figure \ref{fig2}. This implementation involves an exchange of the two bits $a_x,b_z$. If Alice and Bob do not send her/his bit ($a_x$/$b_z$) simultaneously, there would leak information about her/his state. The analysis is referred to Ref.\cite{dupuis2010}. Anyway, it is not secure to directly use this implementation of NL-CNOT.

Based on the construction in Figure \ref{fig2}, we should consider how to securely implement the computation of NL-CNOT. It is obvious from the figure that, there are two nonlocal elements in this circuit: 1) the preparation of the entanglement state $|\chi\rangle$; 2) the remote controlled operation in the dotted rectangle.
The entanglement state $|\chi\rangle$ can be prepared by any one party. After one party has prepared the 4-qubit state, he sends two qubits to the other party. It is secure in the passive adversaries model, because no additional information is revealed in the fixed state.
In the part of the dotted rectangle, Alice's local operation $Z$ is controlled by Bob's measurement result $b_z$, and Bob's local operation $X$ is controlled by Alice's measurement result $a_x$. So, in order to remotely implement NL-CNOT. Alice and Bob must exchange their measurement results $a_x$ and $b_z$.

Above all, in the passive adversaries model, the security of two-party quantum computation of NL-CNOT depends on the security of exchanging two bits ($a_x$ and $b_z$). It must avoid the case: one party has received a bit before sending his bit to the other party. The reason was analyzed in Ref.\cite{dupuis2010}.

Denote $SWAP(a, b)$ the exchange of two bits $a,b\in\{0,1\}$. It can be securely implemented using bit commitment as follows: (1)Alice commits a bit $a$ to Bob, and Bob commits a bit $b$ to Alice; (2)Alice opens her commitment
$a$, and Bob opens his commitment $b$.

Next, the detail is presented about the secure two-party quantum computation of NL-CNOT. Suppose Alice posses the control qubit $|\alpha\rangle$, and Bob posses the target qubit $|\beta\rangle$, and the entanglement state $|\chi\rangle$ is prepared by Alice.
\begin{enumerate}
\item{}
Alice prepares a 4-qubit entanglement state $|\chi\rangle$ (the four qubits are called the 1st, 2nd, 3rd, 4th qubit separately). She sends the 3rd and 4th qubits to Bob, and retains the 1st and 2nd qubits;
\item{}
Alice performs Bell measurement on the control qubit $|\alpha\rangle$ and the 1st qubit, and the result is $a_x,a_z$; Bob performs Bell measurement on the 4th qubit and target qubit $|\beta\rangle$, and the result is $b_x, b_z$;
\item{}
Alice and Bob call the oracle $SWAP(a_x,b_z)$ to exchange the two bits $a_x$ and $b_z$;
\item{}
Alice performs quantum operation $Z^{b_z}X^{a_x}Z^{a_z}$ on the 2nd qubit; Bob performs quantum operation $Z^{b_z}X^{a_x}X^{b_x}$ on the 3rd qubit; The 2nd and 3rd qubits are the output.
\end{enumerate}
It can be verified that, the two-qubit output is $CNOT(|\alpha\rangle\otimes|\beta\rangle)$, and the former qubit belongs to Alice, and the latter belongs to Bob.

In the protocol, the 2nd and 4th steps are local quantum operations. The security of the protocol depends on whether Alice honestly prepares the state $|\chi\rangle$, and whether $SWAP(a_x,b_z)$ is securely implemented. In the passive adversary model, Alice should prepare the state $|\chi\rangle$ honestly following the protocol. Thus, in this model, the security of the protocol depends only on the secure implementation of $SWAP(a_x,b_z)$. Moreover, because $SWAP(a_x,b_z)$ can be implemented based on bit commitment and its security depends on the security of bit commitment, the secure two-party quantum computation of NL-CNOT is reduced to bit commitment.

Bit commitment is a fundamental cryptographic primitive in modern cryptography. It can be implemented using either classical technique or quantum technique. The classical bit commitment is not so secure as quantum bit commitment. Though Refs. \cite{mayers1997,lo1997b} have proved the impossibility of unconditionally secure quantum bit commitment. Later, Refs. \cite{kitaev2004,ariano2007} proved that the impossibility holds under nonrelativistic quantum mechanics. While in the relativistic quantum mechanics, an unconditionally secure quantum bit commitment protocol is proposed by Kent \cite{kent2012}. Thus, this quantum bit commitment scheme can be used to implement secure two-party quantum computation of NL-CNOT.

\section{Multiparty quantum computation scheme and its security}
According to the quantum UC theorem \cite{unruh2010}, in the SMQC protocol proposed in Section \ref{sec3}, the NL-CNOT functionality implemented by TTP can be replaced with the two-party protocol of NL-CNOT proposed in Section \ref{sec4}. Thus, we can obtain a SMQC protocol that does not depend on TTP. Its security is analyzed, and can be concluded briefly: It is secure against passive adversaries, and not secure in the active adversaries model; It can be modified using the technique proposed in Ref. \cite{dupuis2012} to prevent active attack, but there still exists a vulnerability. The detail analysis is as follows.

Next, we analyze the security of the SMQC protocol in the passive adversaries model.

It can be known from the protocol that, all the local qubits belonging to each party have never been transmitted to another party, and the only communication between them is the exchange of classical bits. However, the bits of each party are randomly generated from the local quantum measurements, and leak nothing about the other party's collapsed state after measurements. Then, in the passive adversaries model, no information about the state of another party is revealed, and the dishonest party can gather no information during the computation. Thus, it is secure in the passive adversaries model.

However, it is not secure in the active adversaries model. Next, the analysis is focused on the active attack to the secure computation protocol of NL-CNOT.

In the secure computation protocol of NL-CNOT, when the entanglement state $|\chi\rangle$ is prepared by an active adversary Bob (or Alice), the adversary can perform any Clifford operator on the 2nd (or 3rd) qubit of $|\chi\rangle$, and sends the 1st and 2nd (3rd and 4th) to Alice (or Bob). This causes that the Alice's (or Bob's) result of performing NL-CNOT is transformed with a Clifford operator, which is unknown to Alice (or Bob). Thus, the result is corrupted, however, Alice (or Bob) does not realize the corruption.

In active adversaries model, there are another two kinds of attack strategy to the secure computation protocol of NL-CNOT: (l)Alice can change the measurement basis; (2)Alice flips the measurement bit $a_x$ before exchanging it with Bob's bit $b_z$.

The honest Alice performs the Bell measurement using the basis $\{|B_{xz}\rangle|x,z\in\{0,1\}\}$. However, the dishonest Alice can performs the measurement using another measurement basis, e.g. $\{|B(U)_{xz}\rangle~|~~|B(U)_{xz}\rangle=U^\dagger\otimes I |B_{xz}\rangle, x,z\in\{0,1\}\}$, where $U$ is some unitary transformation. Then, the output of Figure \ref{fig2} is $|out\rangle=CNOT((U|\varphi\rangle)\otimes|\phi\rangle)$, where $|\varphi\rangle$ and $|\phi\rangle$ are the inputs. The details are omitted. The readers can deduce it from the result of teleportation-based quantum computation \cite{jozsa2005}.

Suppose Alice's measurement result is $a_x,a_z$. The dishonest Alice exchanges $a_x'=1-a_x$ with Bob's measurement bit $b_z$. Then the output of Figure \ref{fig2} is $|out\rangle=CNOT(|\varphi\rangle\otimes X|\phi\rangle)$, where $|\varphi\rangle$ and $|\phi\rangle$ are the inputs.

Similarly, the dishonest Bob can also adopt the above two kinds of attack strategy in the active adversaries model.

Above all, the SMQC protocol presented here cannot prevent active adversaries. However, it can be modified using the technique proposed in Ref. \cite{dupuis2012} to prevent active adversaries. Even if the SMQC protocol is transformed into an actively secure protocol, a vulnerability still exists. The analysis is as follows.

Suppose there exists an ideal two-party protocol for NL-CNOT (e.g there exits an ideal black-box that implements NL-CNOT functionality). If Alice and Bob's inputs are $|\varphi\rangle,|\phi\rangle$, where $|\phi\rangle$ is the eigenvector of $X$ ($|+\rangle$ or $|-\rangle$), Alice has perfect attack strategy in the active adversaries model.

{\bf Proposition 1:} Let $|\phi\rangle$ being the eigien state ($|+\rangle$ or $|-\rangle$) of Pauli $X$ operator.
Then for any two qubits $|\varphi\rangle$ and $|\varphi'\rangle$,
$$tr_1[CNOT(|\varphi\rangle\otimes|\phi\rangle)]=tr_1[CNOT(|\varphi'\rangle\otimes|\phi\rangle)],$$
where $tr_1(\cdot)$ denotes the partial trace on the first qubit.

The proof is omitted here. From the proposition, when $|\phi\rangle=|+\rangle$ (or $|-\rangle$), $CNOT(|\varphi\rangle\otimes|\phi\rangle)$ and	
$CNOT(|\varphi'\rangle\otimes|\phi\rangle)$ have the Schmidt decomposition as follows:
$$CNOT(|\varphi\rangle\otimes|\phi\rangle)=\sum_k a_k|\alpha_k\rangle\otimes|\beta_k\rangle,$$
and
$$CNOT(|\varphi'\rangle\otimes|\phi\rangle)=\sum_k a_k|\alpha_k'\rangle\otimes|\beta_k\rangle.$$
Because there exists a unitary transformation $U_1$ that transforms $|\alpha_k\rangle$ to $|\alpha_k'\rangle$, we can conclude that $\exists U_1$, such that
$$CNOT(|\varphi'\rangle\otimes|\phi\rangle)=(U_1\otimes I)CNOT(|\varphi\rangle\otimes|\phi\rangle).$$
This allows Alice to evaluate CNOT between an arbitrary state $|\varphi'\rangle$ and the state $|+\rangle$ (or $|-\rangle$). Moreover, Alice can perform local transformation $U_1$ to get the result of CNOT evaluation on any desired input $|\varphi'\rangle$, but Bob cannot find out Alice's dishonest behavior.

It should be noticed that, although a vulnerability exists in the scheme, it is still a secure scheme. The reason is similar as that, in modern cryptography, the existence of weak key does not mean insecurity of the encryption scheme. Moreover, the input state of Bob is unnecessary to be $|+\rangle$ or $|-\rangle$, and Alice cannot know whether Bob would choose one of the two state as his input.

\section{Discussions}
Ref. \cite{dupuis2010} presented a secure two-party quantum computation scheme of unitaries. Our scheme without TTP is a simplify and extension of theirs. Actually, if the NL-CNOT gate in their scheme is replaced with the protocol proposed in Section \ref{sec4}, their scheme can be simplified into the two-party case of our SMQC protocol. In Ref. \cite{dupuis2010}, the key-updating is the core of their protocol. Because key-updating is relative to the different quantum gates (e.g. Pauli gates, $H, P, T$, local CNOT), and the implementation of $T$ gate needs the call of ideal AND-Box functionality. While in this simplified scheme, key-updating and ideal AND-Box functionality are not necessary, and secure bit commitment is sufficient.

This paper studies SMQC, and shows that bit commitment is universal for SMQC without nonlocal measurements. However, the case that there exists nonlocal quantum measurements in the quantum circuit has not been considered. So it is left an open question: is bit commitment sufficient for secure nonlocal quantum measurement? If it is, then any multiparty quantum computation can be securely implemented based on bit commitment.

According to Ref. \cite{dupuis2012}, the SMQC protocol here can be improved to resist active attack. However, a vulnerability still exists in the active adversaries model. The analysis is shown in the previous section. In consideration of the fundamental role of CNOT in the quantum computation, we conjecture that any SMQC protocol has the vulnerability in the active adversaries model.

\section{Conclusions}
This paper studies SMQC without nonlocal measurements. A SMQC protocol is constructed based on NL-CNOT functionality, which is implemented with TTP. Moreover, in the passive adversary model, a secure two-party quantum computation protocol of NL-CNOT is proposed based on bit commitment. Thus it can be inferred that bit commitment is universal for SMQC without nonlocal measurements. Based on the two-party protocol of NL-CNOT, SMQC protocol can be constructed without TTP. In addition, the security of the two-party protocol of NL-CNOT is analyzed in active adversaries model.


\begin{thebibliography}{00}
%\softraggedright
%\itemsep=-4pt plus.2pt minus.2pt  %% sets the vertical space between items
%\small
\bibitem{lo1997a}
H. K. Lo,  Insecurity of quantum secure computations.  Phys. Rev. A 56(2): 1154-1162, 1997.

\bibitem{colbeck2007}
R. Colbeck, Impossibility of secure two-party classical computation. Phys. Rev. A 76(6): 062308, 2007.

\bibitem{buhrman2012}
H. Buhrman, M. Christandl, and C. Schaffner, Complete insecurity of quantum protocols for classical two-party computation. Phys. Rev. Lett. 109(16): 160501, 2012.

\bibitem{crepeau2002}
C. Cr\'{e}peau, D. Gottesman, and A. Smith,  Secure multi-party quantum computation. FOCS 2002.

\bibitem{benor2006}
M. Ben-or, C. Cr\'{e}peau, D. Gottesman, A. Hassidim, and A. Smith, Secure multiparty quantum computation with (only) a strict honest majority. FOCS 2006.

\bibitem{unruh2010}
D. Unruh, Universally Composable Quantum Multi-party Computation. Advances in Cryptology - EUROCRYPT 2010.

\bibitem{dupuis2010}
F. Dupuis, J. B. Nielsen. and L. Salvail, Secure two-party quantum evaluation of unitaries against specious adversaries. Crypto 2010.

\bibitem{dupuis2012}
F. Dupuis, J. B. Nielsen, and L. Salvail,  Actively secure two-party evaluation of any quantum operation. Crypto 2012.

\bibitem{liang2013}
M. Liang, Symmetric quantum fully homomorphic encryption with perfect security. arXiv: 1304.5087. To appear in Quantum Inf. Process.

\bibitem{goldreich2008}
O. Goldreich, Computational complexity: a conceptual perspective, Cambridge university press. Cambridge, 2008.

\bibitem{gottesman1999}
D. Gottesman, I. L. Chuang, Quantum teleportation is a universal computational primitive. Nature 402, 390-393, 1999.

\bibitem{mayers1997}
D. Mayers, Unconditionally Secure Quantum Bit Commitment is Impossible. Phys. Rev. Lett. 78: 3414-3417, 1997.

\bibitem{lo1997b}
H. K. Lo and H. F. Chau, Is Quantum Bit Commitment Really Possible?  Phys. Rev. Lett. 78: 3410, 1997.

\bibitem{kitaev2004}
A. Kitaev, D. Mayers, and J. Preskill, Superselection rules and quantum protocols. Phys. Rev. A 69, 052326 (2004).

\bibitem{ariano2007}
G. M. D'Ariano, D. Kretschmann, D. Schlingemann, and R. F. Werner, Reexamination of quantum bit commitment: The possible and the impossible. Phys. Rev. A 76(3), 032328 (2007).

\bibitem{kent2012}
A. Kent, Unconditionally secure bit commitment by transmitting measurement outcomes. Phys. Rev. Lett. 109: 130501, 2012.

\bibitem{jozsa2005}
R. Jozsa, An introduction to measurement based quantum computation. arXiv: quant-ph/0508124.

\end{thebibliography}
\end{document}